\documentclass[prl, twocolumn]{revtex4}
\usepackage{graphicx}
\usepackage{epstopdf}
\usepackage{amsmath, amssymb}
\usepackage{color}
\usepackage{xspace}
\usepackage{subfig}

\def \e { \mbox{$\mathrm{e}$} }
\def \vth {v_{\mbox{\scriptsize{th}}}}
\def \vcut {v_{\mbox{\scriptsize{cut}}}}
\def \i {\dot{\imath}}
\def \W {W}

\def \ZNS {Z_{\mbox{\scriptsize{NS}}}}
\def \ENS {E_{\mbox{\scriptsize{NS}}}}

\newcommand\eg{\textit{e.g.}\xspace}
\newcommand\ie{\textit{i.e.}\xspace}

\begin{document}

\title{Energy transfer and dual cascade in kinetic magnetized plasma turbulence}
\author{G. G. Plunk}
\email{gplunk@umd.edu}
\author{T. Tatsuno}
\affiliation{Department of Physics and IREAP, University of Maryland, College Park, Maryland 20742, USA}

\begin{abstract}
The question of how nonlinear interactions redistribute the energy of fluctuations across available degrees of freedom is of fundamental importance in the study of turbulence and transport in magnetized weakly collisional plasmas, ranging from space settings to fusion devices.  In this letter, we present a theory for the dual cascade found in such plasmas, which predicts a range of new behavior that distinguishes this cascade from that of neutral fluid turbulence.  These phenomena are explained in terms of the constrained nature of spectral transfer in nonlinear gyrokinetics.  Accompanying this theory are the first observations of these phenomena, obtained via direct numerical simulations using the gyrokinetic code {\tt AstroGK}.  The basic mechanisms that are found provide a framework for understanding the turbulent energy transfer that couples scales both locally and non-locally.
\end{abstract}

\maketitle

\label{intro-sec}

It is known that a dual cascade occurs in 2D neutral fluid turbulence \cite{kraichnan1967} and also in some reduced models of magnetized plasma turbulence such as the Hasegawa--Mima equation \cite{hasegawa}.  This cascade has the intriguing feature that two invariants are spectrally transfered in opposite senses; that is, the enstrophy flows to small scales while the energy is organized at large scales.  This behavior is due to the fact that the redistribution of these two invariants is constrained by the relationship between their spectral densities, as first explained by Fj{\o}rtoft \cite{fjortoft}; also see \cite{terry-newman, nazarenko-quinn} for application relevant to plasmas.

In gyrokinetics, the turbulent field is a scalar distribution function $g$ \footnote{Formally, $g$ is the gyro-averaged perturbed distribution function; see \cite{plunk-jfm}.}, which varies over both position and velocity space, \ie the particle phase space.  There are two important quadratic quantities associated with this description, namely the free energy $\W$ and the electrostatic energy $E$, which are ``nonlinear invariants'' (that is, conserved by the sole action of the nonlinearity) in {\em both} 2D and 3D.  These quantities represent the energy of turbulent fluctuations.  To study their transfer, we focus on a minimal form of gyrokinetics which retains nonlinear dynamics, but neglects linear drive and damping mechanisms: the homogeneous 2D gyrokinetic equation \cite{plunk-jfm}.

It was theoretically predicted \cite{schek-ppcf, plunk-jfm} and numerically observed \cite{tatsuno-prl} that 2D gyrokinetics has a cascade which exhibits the forward transfer of free energy to smaller scales in {\em both} position and velocity space, while electrostatic energy is transfered inversely to larger scales.  In this letter, we find that this conventional dual cascade is only one possible behavior.  Adapting the arguments of Fj{\o}rtoft, a full range of behavior can be understood by considering how the relationship between the nonlinear invariants constrain evolution of the phase-space spectrum.  This analysis reveals that the direction and locality of the nonlinear transfer actually depend on the way in which the free energy is initially distributed (or injected) across the phase space spectrum.  For certain initial conditions, we find that a dramatic non-local inverse transfer of electrostatic energy will occur, whereby large scales are driven directly by small scales.  In the other extreme, we find that the inverse transfer is replaced by a forward transfer which can be nonlocal or local.  Where the transfer is nonlocal, this provides a mechanism for the damping of large scales by small scales, which could be exploited to reduce the turbulent transport that limits the performance of fusion devices.

\label{defs-sec}

In 2D, the gyrokinetic distribution function $g$ depends on the gyro-center position ${\bf R} = \hat{\bf x}R_x + \hat{\bf y}R_y$, the perpendicular velocity $v$ and the time $t$.  This field evolves nonlinearly via the ${\bf E}\times {\bf B}$ velocity determined from the electrostatic potential $\varphi$ which is a function of position {\bf r} and time $t$ (\ie, see Eqn.~(2.5) of \cite{plunk-jfm}).

A crucial assumption at the outset is the existence of an external scale in velocity space, which we take to be the thermal velocity $\vth$.  This determines the characteristic spatial scale, the (thermal) Larmor radius $\rho = \vth/\Omega$.  We assume that $g(v)$ is attenuated above $\vth$.  In fact, for the perturbed entropy to be a finite quantity, $g(v)$ must fall off at least as fast as a gaussian as $v \rightarrow \infty$.  

We normalize velocity and spatial quantities to the thermal velocity and Larmor radius, respectively (see \cite{plunk-jfm} for the full normalization).  In what follows we will focus on interactions among ``sub-Larmor'' scales, \ie spatial scales and velocity scales that are smaller than the Larmor radius and thermal velocity, respectively.    This is distinct from the typical fluid limit $k \ll 1$, where the formulation of the energetics depends on the reduced description specific to the particular problem of interest.

\label{spectral-decomp-sec}

To obtain a discrete spectral representation, we assume the system to have finite extent in both position and velocity space.  Adapting the continuous representation from \cite{plunk-jfm}, we decompose $g$ using a Bessel-Fourier series:

\begin{equation}
g({\bf R}, v) = \displaystyle\sum_{{\bf k_i},p_j}\e^{\i {\bf k}_i\cdot{\bf R}} \frac{\pi p_j}{\vcut} \Theta(\vcut-v)J_0(p_j v)\hat{g}({\bf k}_i, p_j)
\end{equation}

\noindent where $p_j = \lambda_j/\vcut$ and $\lambda_j$ is the $j\mbox{th}$ zero of the zeroth order Bessel function $J_0$ and $\Theta$ is the Heaviside step function, which explicitly provides the velocity cutoff.  The cutoff velocity is $\vcut(k_i) \gg 1$, chosen for each $k_i$ such that $p_j = k_i$ for some $j$.  Inverting the transform gives

\begin{equation}
\hat{g}({\bf k}_i, p_j) = \int \frac{d^2{\bf R}}{L^2}\e^{-\i {\bf k}_i\cdot{\bf R}} \int v dv J_0(p_j) g({\bf R}, v)
\end{equation}

\noindent  where $L$ is the system size in $R_x$ and $R_y$.  In spectral form, the quasi-neutrality constraint, which relates the potential to the distribution function, is simply

\begin{equation}
\hat{\varphi}({\bf k}_i) = \beta({\bf k}_i) \hat{g}({\bf k}_i, k_i)\label{qn-g-k}
\end{equation}

\noindent where $\beta = 2\pi/(1 + \tau - \hat{\Gamma}_0)$, $\tau$ is a constant related to the electron response \cite{plunk-jfm} and $\hat{\Gamma}_0 = I_0(k^2)e^{-k^2}$ where $I_0$ is the zeroth-order modified Bessel function.  For $k \gg 1$, the regime of interest, $\hat{\Gamma}_0 \propto 1/k$ may be neglected.  The electrostatic energy is defined

\begin{equation}
E = \frac{1}{2}\int \frac{d^2{\bf r}}{L^2}\left[(1 + \tau)\varphi^2 - \varphi\Gamma_0\varphi \right] = \sum_{\bf k_i}E({\bf k}_i)\label{E-def}
\end{equation}

\noindent where, from Eqn.~\ref{qn-g-k}, $E({\bf k}_i) = \pi \beta |\hat{g}({\bf k}_i, k_i)|^2$.  The other invariant, being a function of velocity space, we call ``kinetic free energy'': $G(v) = 1/(2L^2)\int d^2{\bf R}\; g^2$.  For the purposes of this paper, we are not interested in this full invariant but only on the consequences its existence has for constraining the spectral evolution of fluctuations (note that \cite{zhu-hammett} do retain the details of $G$).  The relevant quantity is formed by integrating $G$ over its velocity-space dependence; henceforth, we focus on this quantity, the (generalized) free energy:

\begin{equation}
\W = \frac{4\pi\vcut}{1+\tau} \int v dv \; G(v) =  \displaystyle\sum_{\bf k_i}\sum_{p_j} \; W(k_i, p_j).\label{Wg-def}
\end{equation}

\noindent where $\W({\bf k}_i, p_j) = 2\pi^2 p_j |g({\bf k}_i, p_j)|^2/(1+\tau)$.  For comparison we note that $\W$ is proportional to the quantity $W_{g1}$ from \cite{plunk-jfm}.  Note that $\W$ (or more generally $G(v)$) and $E$, being the known quadratic invariants, play a special role in constraining spectral transfer, just as enstrophy and energy do in 2D fluid turbulence \footnote{The perturbed entropy is not an independent quantity as it can be written in terms of $\W$ and $E$ (see \cite{plunk-jfm}).}.

\label{constrained-sec}

We now arrive at the central equation of this letter, a constrains on the spectral densities of the invariants.  In the limit $k \gg 1$, it follows from the above definitions that (see also the derivation \cite{plunk-jfm}, Eqn. (7.14))

\begin{equation}
\W({\bf k}_i, k_i) = k_iE({\bf k}_i).\label{E-W-spec-reln}
\end{equation}

\noindent Upon initial inspection, this relationship is similar to that for the NS equation: $\ZNS({\bf k}_i) = k_i^2 \ENS({\bf k}_i)$, where $\ENS$ and $\ZNS$ are the energy and enstrophy, respectively.  However, this relationship constrains energetic interactions involving those components of free energy residing {\em only} along the diagonal in $k$-$p$ space ($k_i = p_j$) \footnote{This constraint is strong since, by Eqn.~\ref{qn-g-k}, excitation of the diagonal is necessary for any nonlinear dynamics; \ie, the nonlinearity is zero if $\varphi$ is zero.}.  This fact underscores the difference between the roles of velocity space and position space and, as we will see, is the central reason for the novel behavior that is discovered.  Let's first consider what Eqn.~\ref{E-W-spec-reln} implies for three-scale interactions.
\label{three-scale-sec}

We define the free energy at the wavenumber pair $(k_i, p_i)$ to be $\W_i = W(k_i, p_i) = \sum_{|{\bf k}_j| = k_i} W({\bf k}_j, p_i)$ and the corresponding electrostatic energy to be $E_i = \delta(k_i - p_i)E(k_i) = \delta(k_i - p_i)\sum_{|{\bf k}_j| = k_i} E({\bf k}_j)$, where $\delta$ is the discrete delta function.  Let's consider the difference in energies between the initial and final times, $t_0$ and $t_1$, for three components $i = 1$, $2$ and $3$: Thus we define $\Delta W_i = W_i(t_1) - W_i(t_0)$ and $\Delta E_i = E_i(t_1) - E_i(t_0)$.  Conservation of $\W$ and $E$ implies:

\begin{equation}\label{WE-conserve}
\begin{split}
&\Delta \W_1 + \Delta \W_2 + \Delta \W_3 = 0,\\
&\Delta E_1 + \Delta E_2 + \Delta E_3 = 0.
\end{split}
\end{equation}

\begin{figure}
\subfloat[I: Fj{\o}rtoft-type]{\label{transitions-fig-a}\includegraphics[width=0.45\columnwidth]{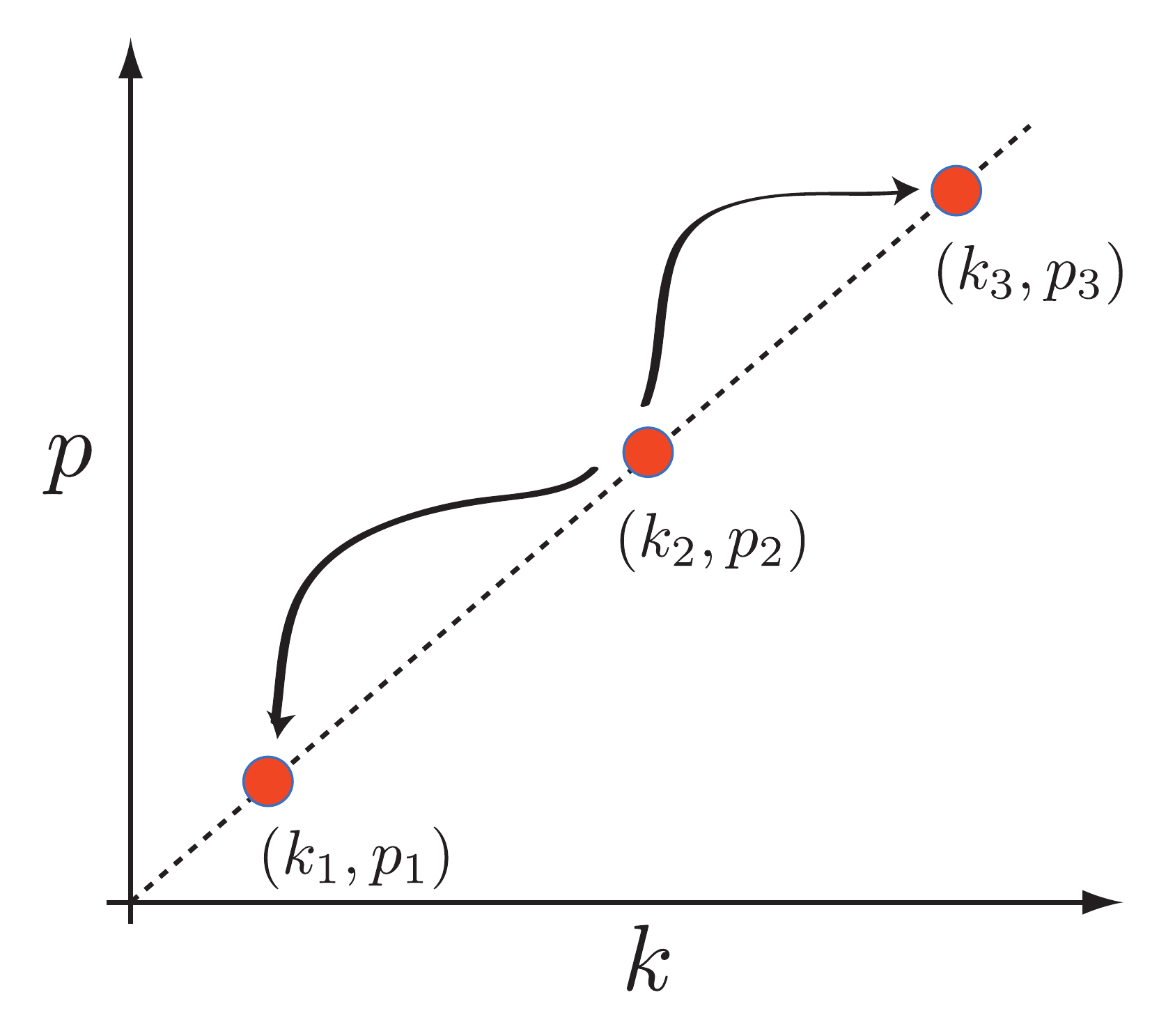}}
\subfloat[II: Kinetic-type]{\label{transitions-fig-b}\includegraphics[width=0.45 \columnwidth]{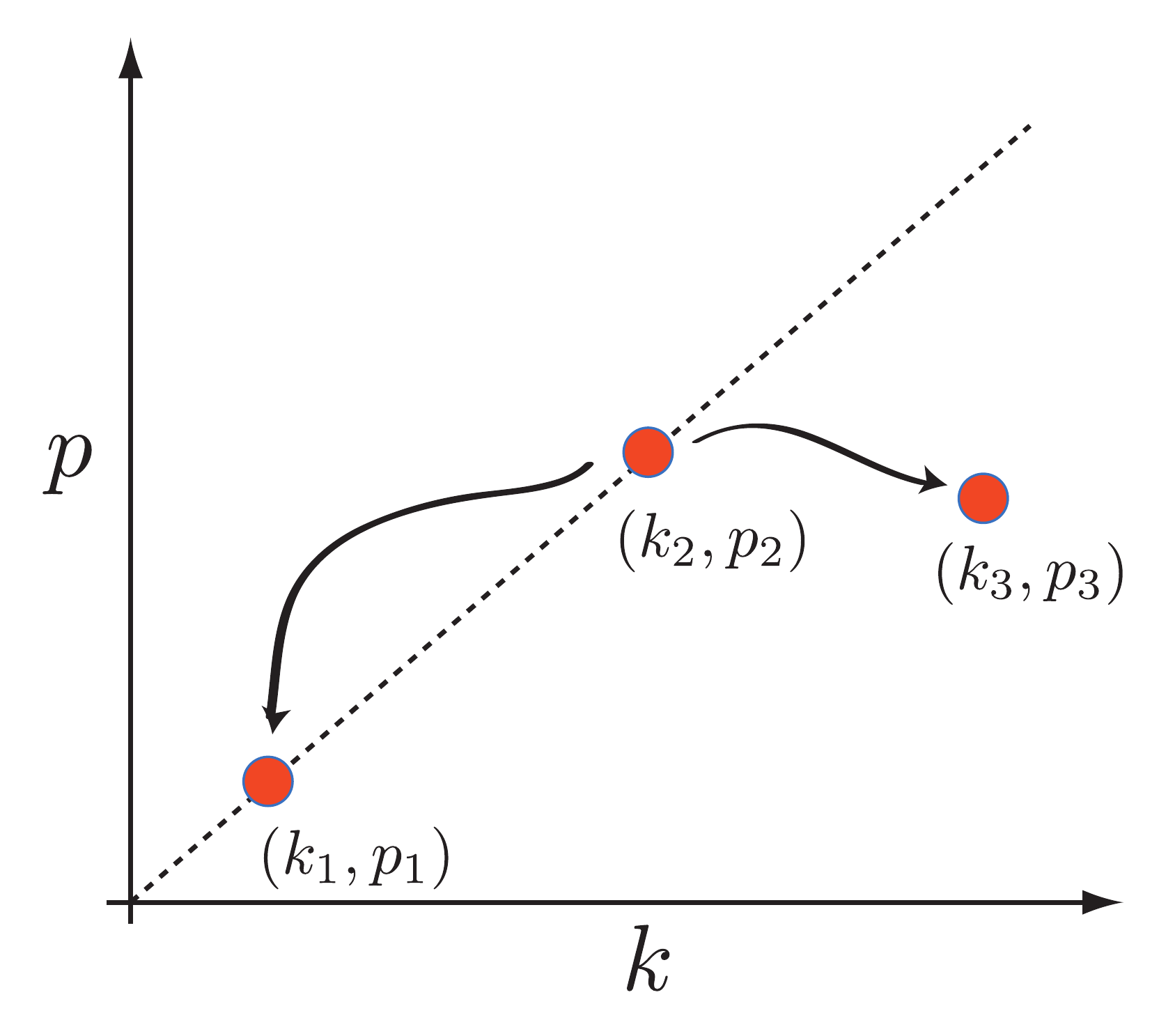}}
\caption{Constrained energetic transitions involving three scales.  The dotted diagonals indicate $k = p$.}
\label{transitions-fig}
\end{figure}

As illustrated in Fig.~\ref{transitions-fig}, there are two types of constrained transfer.  For definiteness, we take $k_1 < k_2 < k_3$.

In type I transitions, Fig.~\ref{transitions-fig-a}, all three components are diagonal, $p_i = k_i$.  In this case, transfers are constrained in precisely the sense of Fj{\o}rtoft, with the substitution $k_i^2 \rightarrow k_i$.  From equations Eqns.~\ref{E-W-spec-reln} and \ref{WE-conserve} we obtain:

\begin{eqnarray}
\Delta E_1 = -\Delta E_2(k_3 - k_2)/(k_3 - k_1),\\
\Delta E_3 = -\Delta E_2(k_2 - k_1)/(k_3 - k_1).
\end{eqnarray}

\noindent The quantities in parentheses are positive (because $k_1 < k_2 < k_3$).  Thus, as noted by Fj{\o}rtoft, it is only the intermediate wavenumber, $k_2$, which can be a source for both the two remaining components.

Type II transitions, Fig.~\ref{transitions-fig-b}, involve two diagonal components, and also lead to inverse transfer of $E$ for the arrow directions indicated.  To see this, first note that $\Delta E_3 = 0$ because $p_3 \neq k_3$.  From Eqns.~\ref{E-W-spec-reln} and \ref{WE-conserve} we find

\begin{equation}
\Delta E_1 = \Delta \W_3/(k_2 - k_1).
\end{equation}

Thus, a transfer of free energy from diagonal to non-diagonal components must be accompanied by a simultaneous inverse transfer of $E$; as we will shortly see, the reverse process can also spontaneously occur.

Finally, note that transitions involving only one diagonal component are forbidden as they cannot conserve $E$, while those that involve no diagonal components are unconstrained.  

\label{examples-sec}

The elementary three-scale transitions form building blocks for thinking about more complicated evolution.  For instance, the forward diffusion of free energy along and away from the diagonal, generally speaking, must induce an inverse transfer of electrostatic energy.  On the other hand, we can imagine that the free energy could be initially concentrated in non-diagonal components and, in this case, the random redistribution of free energy would excite diagonal components causing electrostatic energy to flow forward, in the reverse sense of Fig.~\ref{transitions-fig-b}.  From these considerations, it seems that the tendency for electrostatic energy to flow forward or backward should depend on the way in which the free energy is distributed relative to the diagonal.  We will use the parameter $\kappa = \W/E$ to help quantify this: small $\kappa$ corresponds to free energy being concentrated on the diagonal while large $\kappa$ corresponds to free energy being concentrated off-diagonal.

\begin{figure}
\begin{center}
\subfloat{\label{ex-spectra-fig-a1}}\subfloat{\label{ex-spectra-fig-a2}}\subfloat{\label{ex-spectra-fig-b1}}\subfloat{\label{ex-spectra-fig-b2}}\subfloat{\label{ex-spectra-fig-c1}}\subfloat{\label{ex-spectra-fig-c2}}
\includegraphics[width=\columnwidth]{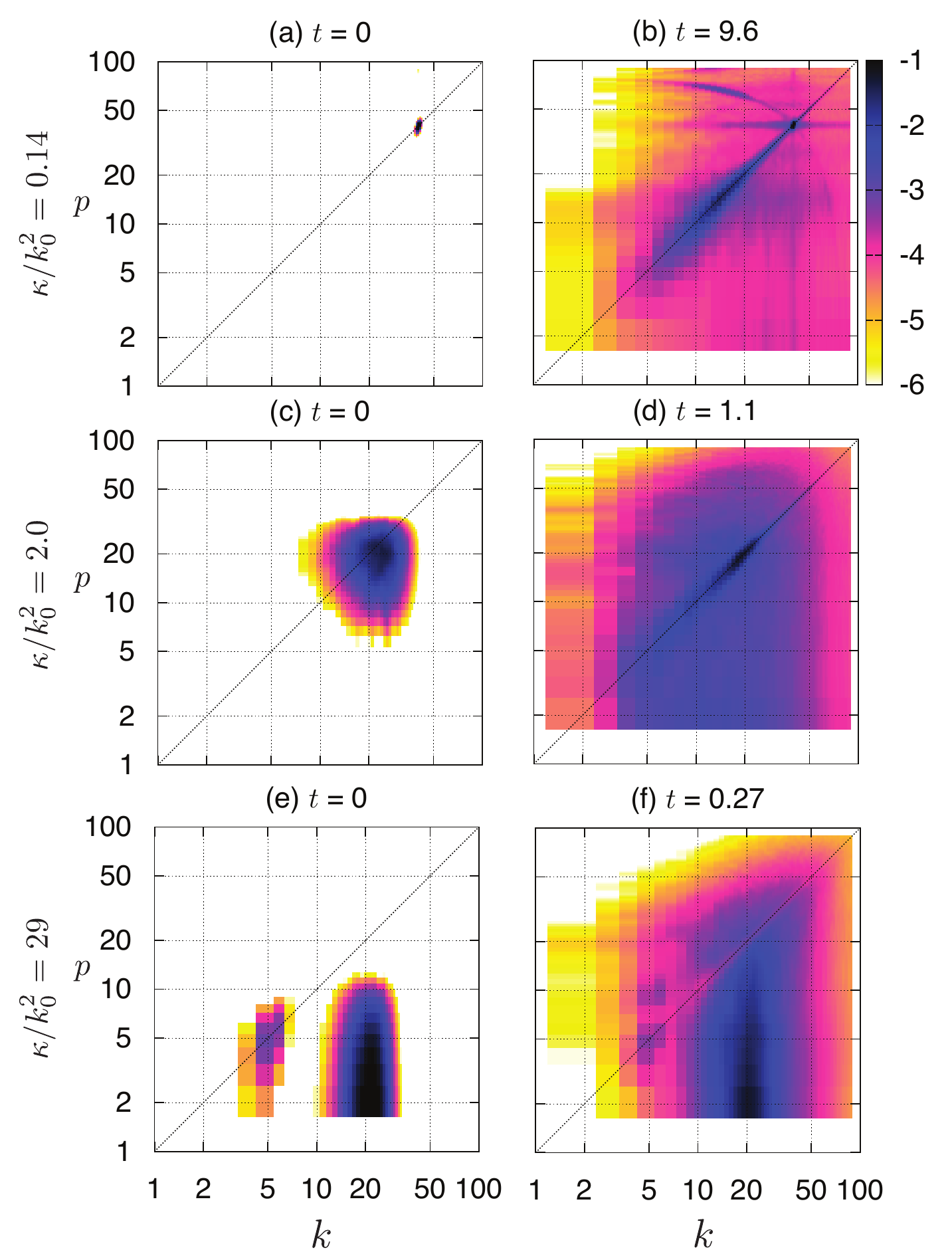}
\caption{Example spectral distributions $\log_{10} [\W(k,p)/\W]$ and initial evolution.  Diagonals marked by dotted lines.}
\label{ex-spectra-fig}
\end{center}
\end{figure}

\begin{figure}
\begin{center}
\subfloat[$\kappa/k_0^2 = 0.14$]{\label{espec-t-fig-a}\includegraphics[width=0.504\columnwidth]{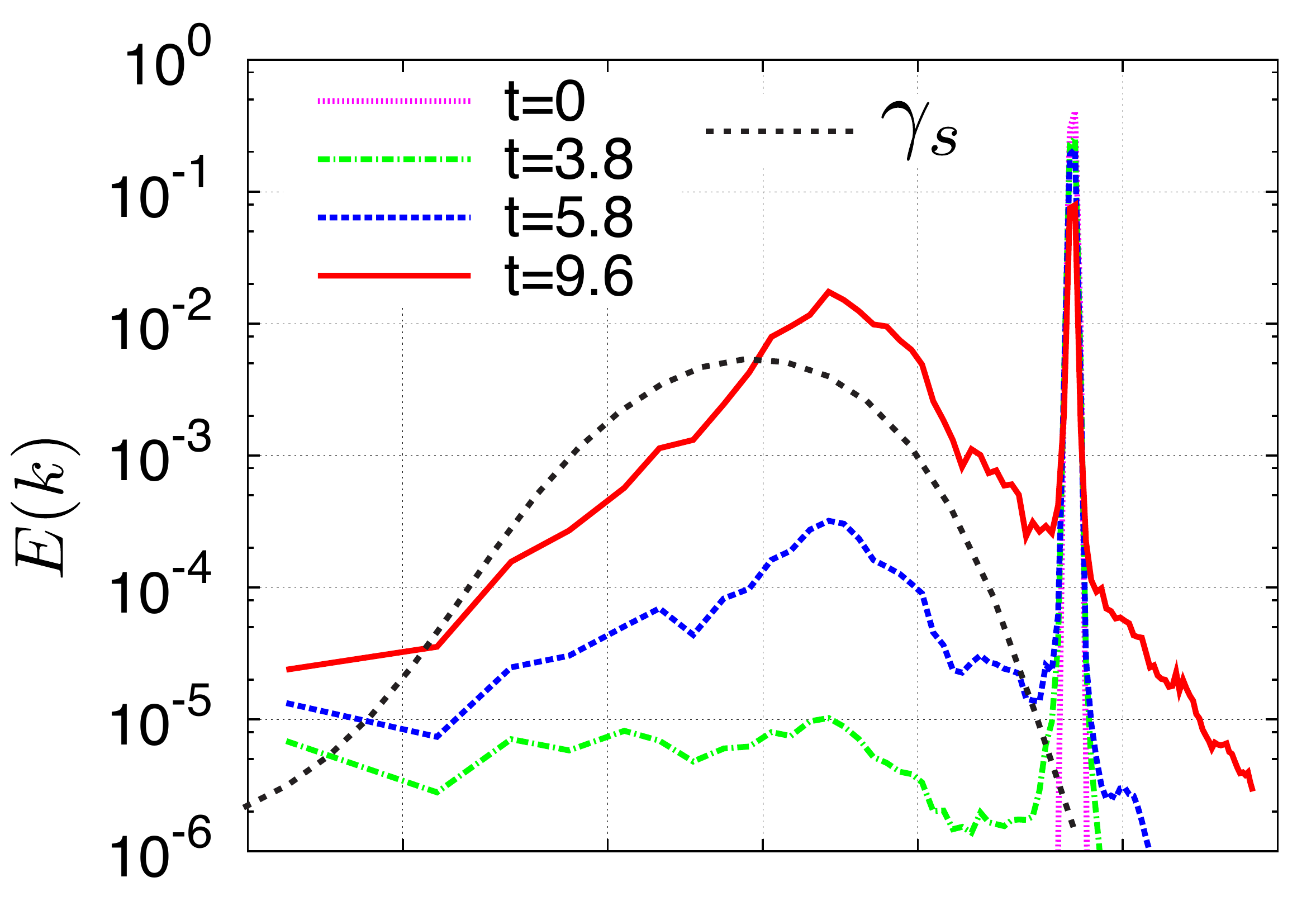}}
\subfloat[$\kappa/k_0^2 = 2.0$]{\label{espec-t-fig-b}\includegraphics[width=0.496\columnwidth]{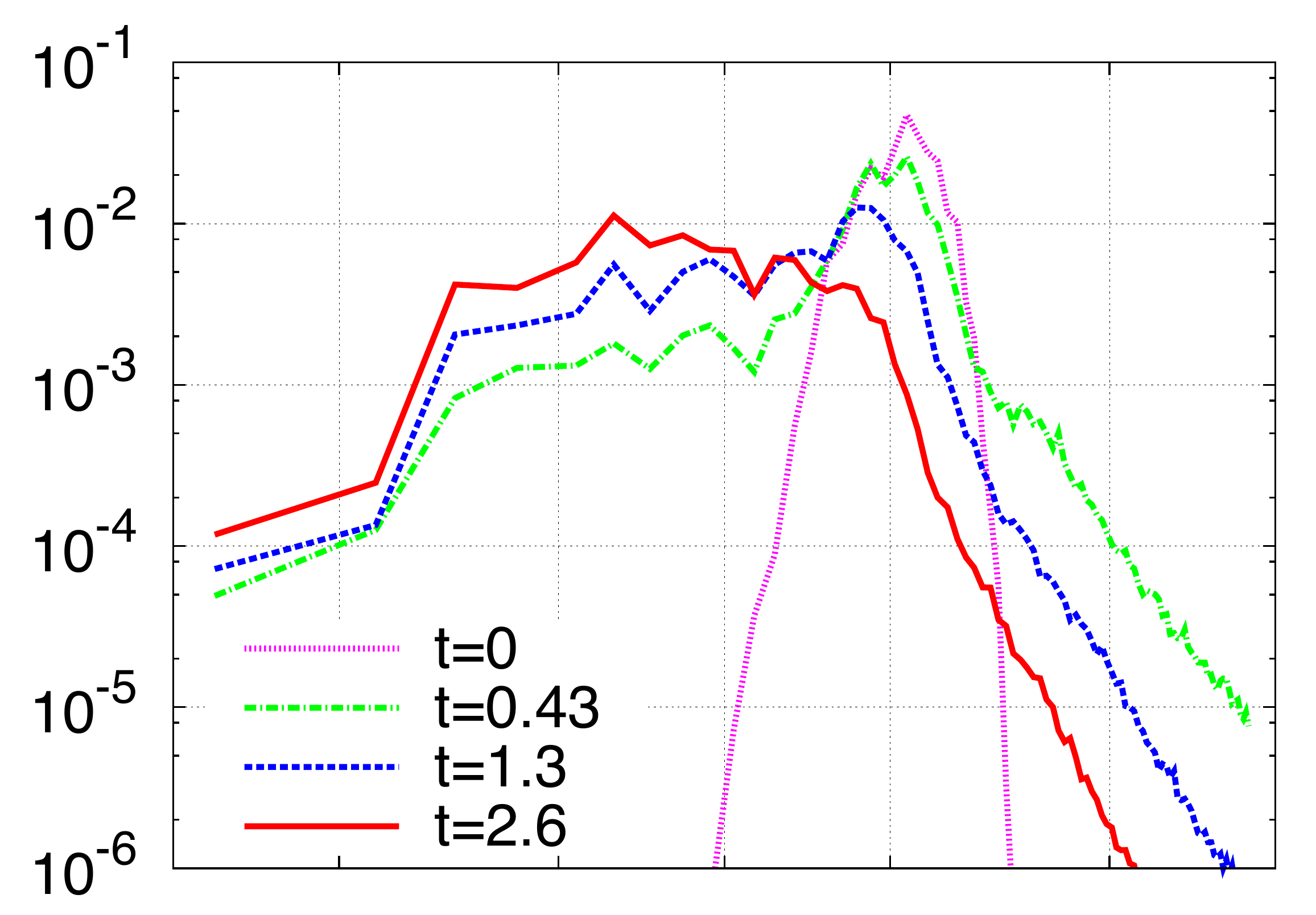}}\\
\subfloat[$\kappa/k_0^2 = 29$, $k_d = 5$, $k_0 = 20$]{\label{espec-t-fig-c}\includegraphics[width=0.505\columnwidth]{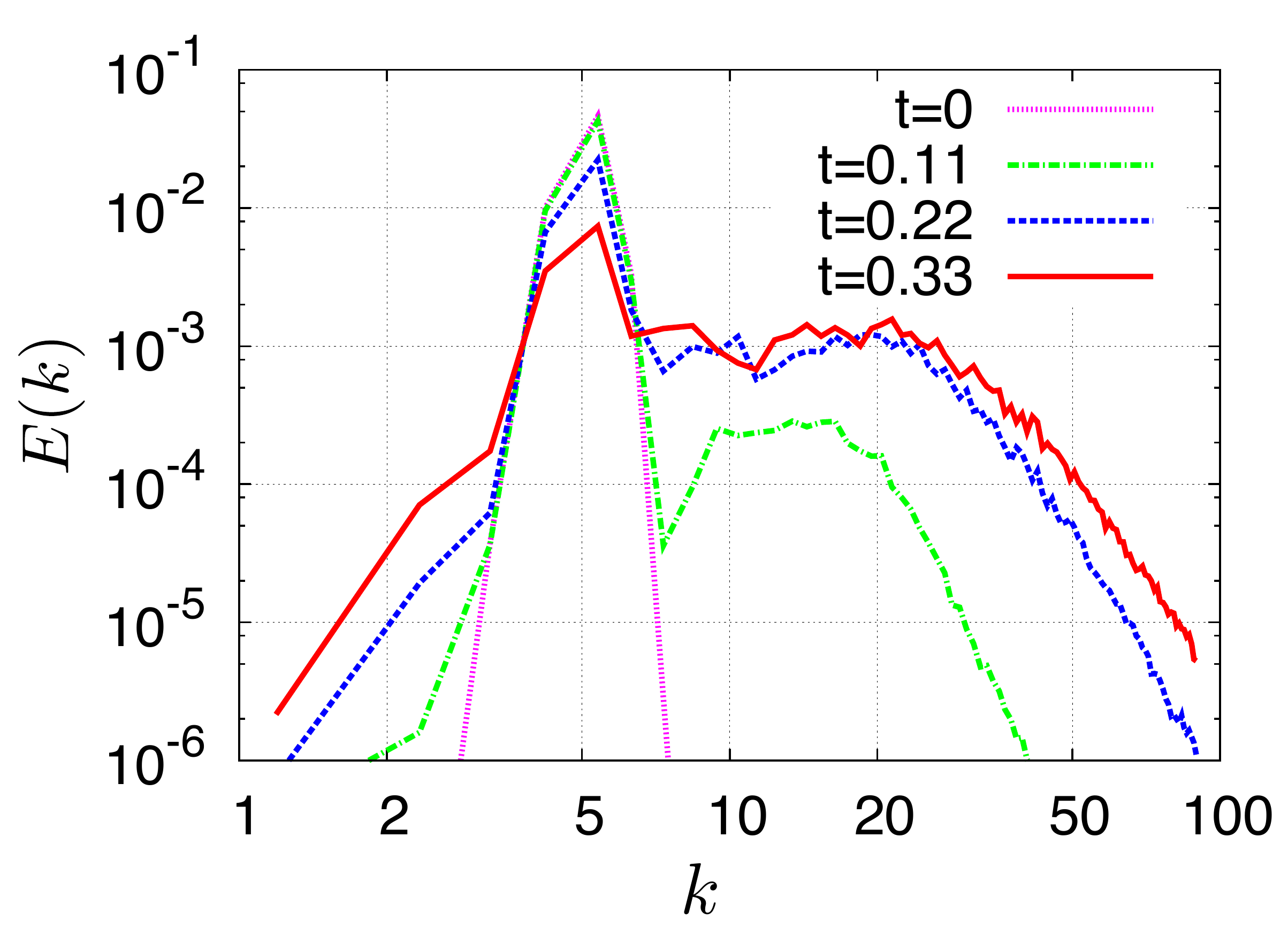}}
\subfloat[$\kappa/k_0^2 = 24$, $k_d = k_0 = 5$]{\label{espec-t-fig-d}\includegraphics[width=0.495\columnwidth]{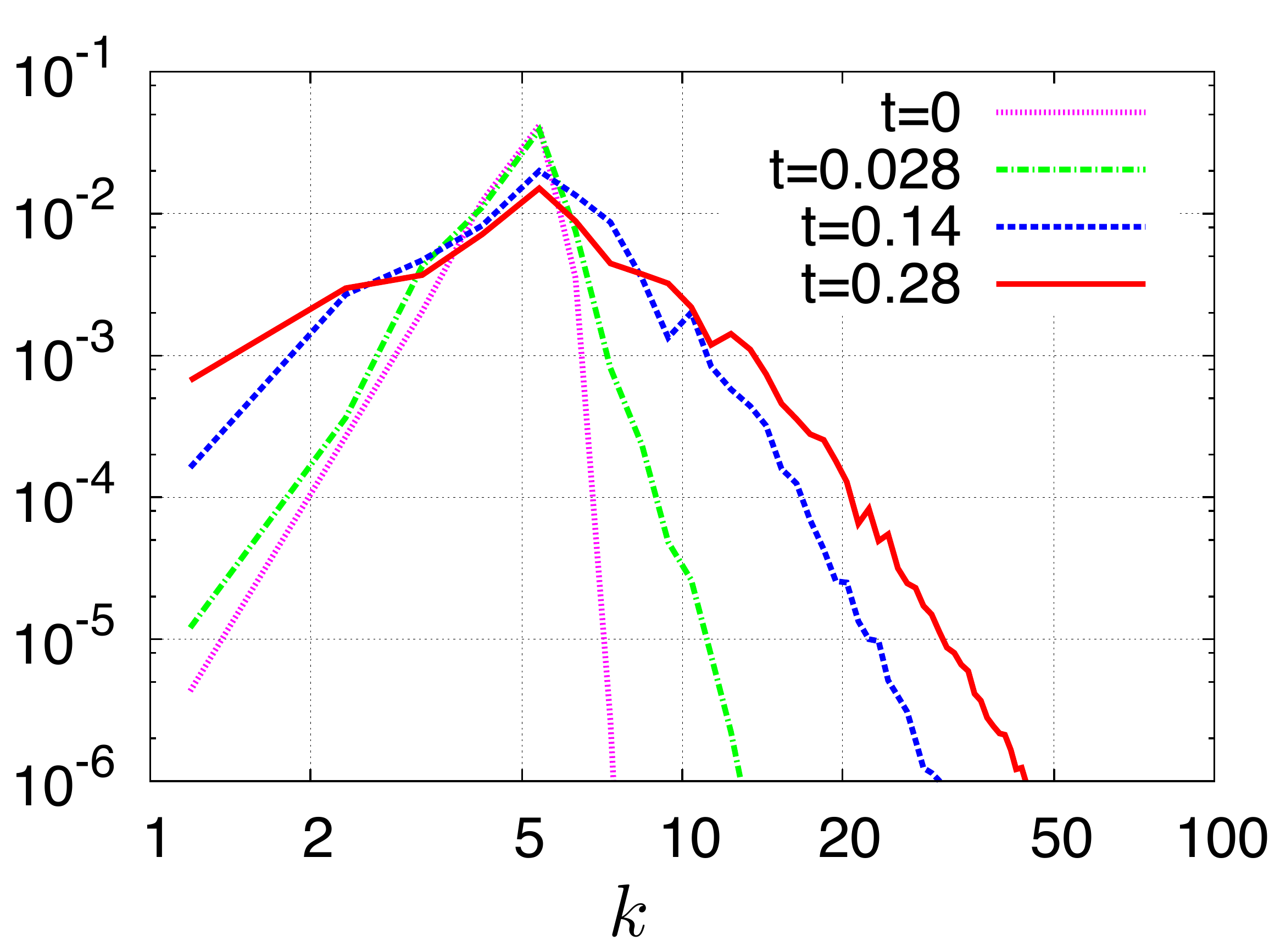}}
\caption{Evolution of the electrostatic energy spectrum.}
\label{espec-t-fig}
\end{center}
\end{figure}

Let us now turn to the numerical results.  We use the nonlinear gyrokinetic code {\tt AstroGK} \cite{numata}, including collisional dissipation as a free energy sink at large $k$ to prevents nonphysical ``bottle-necking''.  We also restrict attention to cases where the lowest wavenumber is unexcited, so as to avoid effects of energetic condensation at large scales.

Having explored a variety of initial spectral distributions, we present three here, corresponding to different magnitudes of $\kappa$, which exemplify different types of spectral evolution.  To accompany the discussion, we provide some simple analytic calculations.  Fig.~\ref{ex-spectra-fig} shows the initial evolution of $\W(k,p)$ for the three cases, in the order of increasing $\kappa$.

The first case, Fig.~\ref{ex-spectra-fig-a1}-\ref{ex-spectra-fig-a2}, is an extreme case, corresponding to small $\kappa$.  The free energy is concentrated around a single diagonal component $(k_0, k_0)$.  The transfer of free energy away from this initial peak can only be accomplished by either Fj{\o}rtoft-type (Fig.~\ref{transitions-fig-a}) or kinetic-type transitions (Fig.~\ref{transitions-fig-b}), which both demand an inverse transfer of electrostatic energy.  As shown in Fig.~\ref{ex-spectra-fig-a2}, this is indeed what occurs: diagonal components are strongly excited at wavenumbers smaller than $k_0$.  As seen in Fig.~\ref{espec-t-fig-a}, the transfer that occurs is actually nonlocal, with a dominant peak of electrostatic energy $E(k)$ emerging spontaneously at a wavenumber, $k_d$, that is significantly smaller than $k_0$. 

Notice in Fig.~\ref{ex-spectra-fig-a2} that the $\W(k, p)$ spectrum is excited along lines in the $k$-$p$ plane.  This hints to the presence of an instability that involves the coupling of a small number of Fourier components.  We confirm that a four-wave truncation (see, \eg, \cite{chen-lin-white}) is in fact an excellent approximation for the instability that is induced in the presence of a single Fourier component; the details of this calculation will be reported elsewhere \footnote{Instability theory (\eg secondary or modulational), as well as absolute statistical equilibrium \cite{zhu-hammett}, give alternative perspectives for studying energy transfer.  We note that although these are not equivalent, they clearly must be consistent with the arguments of Fj{\o}rtoft.}.  The growth rate curve ($\gamma_s(k)$ in arbitrary units) of this instability is incorporated into Fig.~\ref{espec-t-fig-a}.  Note that the wavenumber of peak growth rate is close to that of the peak which develops in the energy spectrum.

At a later stage, the case of Fig.~\ref{ex-spectra-fig-a1} evolves to a state in which the free energy is more broadly distributed about the diagonal.  In this state, the 4-wave instability is apparently suppressed, or otherwise lost in the bath of fluctuations.  Thus, many more degrees of freedom participate and the evolution becomes much more cascade-like, whereby the free energy diffuses locally in $k$-$p$ space.

To treat the cases of Figs.~\ref{ex-spectra-fig-b1}-\ref{ex-spectra-fig-c2} more quantitatively we perform a calculation, extending the arguments of Fj{\o}rtoft with some simple assumptions.  In each case that we consider, the initial spectrum is dominated by a single wavenumber $k_0$.  During a finite period of time, energy is liberated from this distribution in the amount $\W_0$ and $E_0$, satisfying $\W_0/E_0 = \kappa$.  Some amount of these energies, $\W^{\prime}$ and $E^{\prime}$, diffuse forward in $k$-$p$ space.  We observe that this redistribution is locally diffusive in $k$-$p$ space, spreading randomly to unoccupied wavenumbers.  By simple mode counting, this implies the relationship $\W^{\prime} \approx Ck_0^2 E^{\prime}$, where $C$ is some (possibly universal) constant of order unity.  Now, in order to maintain the balance of both invariants, there must be some transfer of electrostatic energy $E_d$ to a wavenumber $k_d$, along with the corresponding free energy $\W_d = k_d E_d$.  By conservation of energies we have $\W_0 = \W^{\prime} + \W_d$ and $E_0 = E^{\prime} + E_d$.  We can combine these expressions to obtain

\begin{equation}
R = (\kappa - k_d)/(Ck_0^2 - \kappa)\label{es-energy-ratio},
\end{equation}

\noindent where $R = E^{\prime}/E_d$.  The parameters $R$ and $k_d$ characterize the direction and locality of the transfer of $E$.  From Eqn.~\ref{es-energy-ratio}, it is clear that $\kappa$ is an important parameter: at the critical case $\kappa_c = Ck_0^2$, $R$ diverges and changes sign.

In Fig.~\ref{ex-spectra-fig-b1}, we have an initial condition with free energy distributed broadly around $k_0 = 20$.  This case is the closest in behavior to the conventional dual cascade of fluid mechanics.  The free energy is observed to diffuse locally forward in $k$-$p$ space whereas the transfer of electrostatic energy is mostly local and directed in the inverse direction, to $k < k_0$, as is apparent in Fig.~\ref{espec-t-fig-b}.  Thus it appears that $0 < R < 1$, so, since $\kappa > k_0 > k_d$, we can conclude that $\kappa < \kappa_c$.

For $\kappa \rightarrow \kappa_c = C k_0^2$, the evolution of $E(k)$ changes character sharply.  If, for instance, the magnitude of $k_d$ is comparable to $k_0$, \ie $k_d \sim {\cal O}(k_0)$, then, by Eqn.~\ref{es-energy-ratio}, $R$ is a large positive number for $\kappa \lesssim \kappa_c$.  Therefore most of the electrostatic energy transfered is forward.  Now, if $\kappa$ exceeds $\kappa_c$, $R$ becomes negative and electrostatic energy is extracted from the $k_d$ component.  For $k_d > k_0$, this then corresponds to an inverse transfer of $E$, while for $k_d < k_0$, the transfer is forward.

In Figs.~\ref{ex-spectra-fig-c1}-\ref{ex-spectra-fig-c2} we have the case $\kappa \gg \kappa_c$.  Here there is a dominant bath of free energy around $k \sim 20$ while we fix $k_d$ at the outset by providing an initial ``seed'' energy at $k_d \sim 5$.  As the system evolves, electrostatic energy is drawn from the component at $k_d$ and so, by Eqn.~\ref{es-energy-ratio}, $E_d$ is negative.  This process must proceed by kinetic transitions in the reverse sense of Fig.~\ref{transitions-fig-b}, whereby electrostatic energy is ``forcefully'' drawn from the large scale components to support the diffusion of free energy at small scales.

The local diffusion of free energy is clearly observed from an initial bath at high-$k$ and low-$p$.  Also, the evolution of $E(k)$, shown in Fig.~\ref{espec-t-fig-c}, shows forward transfer of electrostatic energy.  For comparison, Fig.~\ref{espec-t-fig-d} shows another large-$\kappa$ case, but with the seed energy at $k_d = k_0$.  Here the system evolves by a dual forward cascade, whereby both $\W$ and $E$ are transfered locally to finer scales.  Note that the surprising ``cascade reversal'' of $E$ is essentially due to the freedom afforded by the larger spectral space available to $\W$.  Note that the directionality and locality of transfers in all of the above cases has been confirmed by a nonlinear transfer diagnostic; these observations will be reported elsewhere.

It is true that in a general 3D gyrokinetic system, magnetic geometry and linear mode physics can formally break the invariance of $E$ and $\W$ by introducing both damping and instability; this strongly affects the nature of the turbulence.  However, we stress that the nonlinear transfer of fluctuation energy must always occur in such a way as to conserve both $E$ and $\W$.  This is due to the fact that the nonlinearity can always be written in terms of a Poisson bracket involving a stream function (in the electrostatic limit, the electrostatic potential) for the turbulent flow; this is not possible in 3D fluid turbulence (or general plasma turbulence) and thus energy transfer is not constrained as it is in 2D.  Because of these facts, we believe that the mechanisms revealed in this letter are a generic feature of kinetic magnetized plasmas.

The authors gratefully acknowledge support and advice from Bill Dorland, and stimulating discussions with Jian-Zhou Zhu, Alexander Schekochihin, Michael Barnes and Greg Hammett.  This work was supported by the Leverhulme Trust Network and U.S. DOE Grant No. DESC0005106.

\bibliography{gyrokinetic-fjortoft-analysis}

\end{document}